\documentclass{moriond}

\usepackage{amsfonts}

\def\Journal#1#2#3#4{{#1} {\bf #2}, #3 (#4)}

\def\PRD{{\em Phys. Rev.} D}

\def\JCAP{\em J. Cosmol. Astropart. Phys.}

\def\be{\begin{equation}}
\def\ee{\end{equation}}
\def\bea{\begin{eqnarray}}
\def\eea{\end{eqnarray}}

\newcommand{\Photo}{\includegraphics[height=35mm]{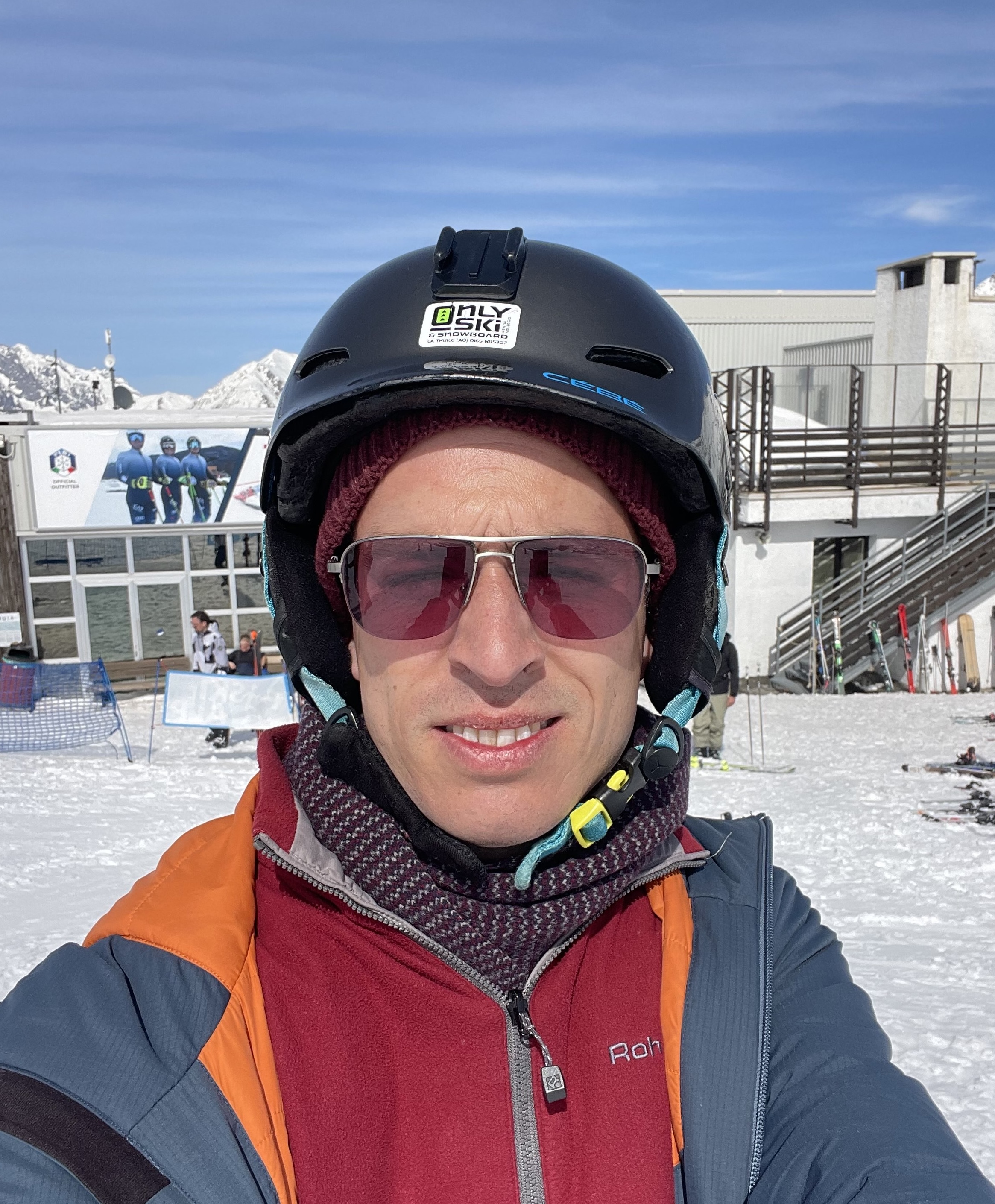}}

\begin{document}
\vspace*{4cm}
\title{A DHOST model of inflation: CMB constraints from the power spectrum and the bispectrum}

\author{Andrei Lazanu}

\address{Department of Physics and Astronomy, University of Manchester, Manchester, M13 9PL, UK}

\maketitle\abstracts{
We build an inflationary model based on Degenerate Higher Order Scalar Tensor (DHOST) theories in a de Sitter background. We determine the scale-dependent power spectrum of curvature perturbations in these theories and we show that such a model can be compatible with the latest \textit{Planck} measurements on the Cosmic Microwave Background (CMB). We calculate the bispectrum of curvature perturbations in DHOST models and we directly constrain them using the \textit{Planck} results on non-Gaussianities. We show that the bispectrum consists of a contribution depending only on the power spectrum parameters and a linear combination of terms depending on new parameters. The former peaks in the squeezed limit, while the latter in the equilateral limit. We use the publicly available CMB-BEST code to directly compare the model predictions to the CMB bispectrum statistics and to marginalise over the free parameters,  explicitly showing that there are viable DHOST inflationary models that satisfy both the power spectrum and bispectrum constraints from \textit{Planck}.}

\section{DHOST inflation}

Degenerate Higher Order Scalar Tensor (DHOST) theories are the most general scalar-tensor theories with one scalar field and which propagate one scalar degree of freedom and no ghosts. These theories can be employed to build inflationary models. In order to keep the theory ghost free and to satisfy certain constraints, we start with a simplification of the DHOST action
\begin{eqnarray}\label{action-dhost}
S_{\rm D} = \int d^4 x \sqrt{-g} \bigg[ f_0(X) + f_1(X) \Box \phi + f_2(X) R
+ \frac{6 f_{2,X}^2}{f_2} \phi^{\nu} \phi_{\nu\eta} \phi^{\eta\lambda}\phi_{\lambda} \bigg]\,,
\end{eqnarray}
where $X=g^{\nu\eta}\phi_{\nu}\phi_{\eta}$ and  $\phi_{\nu}\equiv\nabla_{\nu}\phi$. As such theories yield flat power spectra, which are incompatible with observations, we add perturbations through the potential
\begin{equation}
S_{\rm V} = \int d^4 x \sqrt{-g} \bigg[ - \frac{m^2}{2}  \phi^2 - \frac{\lambda}{4!}  \phi^4 \bigg] \,.    
\end{equation}
In order to determine the power spectrum and the bispectrum of curvature perturbations to compare with \textit{Planck} predictions, we need to expand the action at second and third order respectively, from which the power spectrum and bispectrum in the model can be obtained via field quantisation.

\section{Power spectrum and bispectrum results}
The power spectrum can be determined in terms of four parameters, $\alpha_B$, $\alpha_H$, $\alpha_K$, $\beta_K$, together with the two perturbation parameters~\cite{dhost1}. We obtain values for these parameters using the \textit{Planck} constraints on the amplitude of the scalar power spectrum, its spectral index $n_s$, as well as its running and running of the running and keeping the tensor-to-scalar ratio lower than the current experimental constraints.  Parameter values that satisfy these constraints are given by: $f_2=2.70$, $\alpha_B=1$, $\alpha_H=1.04$, $\beta_K=3.97343$,  $m^2 = -1.6 \times 10^{-23}$, $\lambda =  10^{-36}$ and  $h_{\mathrm{ds}}=3 \times 10^{-5}$, where  $h_{\mathrm{ds}}=\sqrt{-f_0/(6f_2)}$ is the reduced Hubble constant.

In the case of the bispectrum, there are five additional parameters, coming for the second and third order derivatives of the $f_i$'s~\cite{dhost2}. As a first approximation, we consider the bispectrum in  the local, equilateral and orthogonal templates, which have been constrained by \textit{Planck}. One set of such parameters is: $f_{0,{\rm xxx}}=2\times 10^{-6}$, $f_{1,{\rm xxx}} = -0.16$, $f_{2,\rm {xxx}} = 150$, $\beta_B = 0$ and $\beta_H = 0$.

The bispectrum consists of a baseline term, fixed by the power spectrum parameters, and five contributions depending linearly on the new parameters~\cite{dhost3}. The baseline component has a squeezed shape, while the five new components all have indistinguishable equilateral shapes. Thus, the bispectrum can be expressed as $B=v_1B^{\rm baseline}+v_2 B^{\rm new}$, with $B^{\rm new}$ being the linear combination of five contributions and $v_1=1$ corresponding to our model.

To confront our predictions with the \textit{Planck} data temperature and polarisation data, we employ the \textsc{CMB-BEST} code~\cite{cmbbest}, which constrains bispectrum shapes. By running the pipeline, we show that the model originally considered, although providing a good match for the power spectrum, is incompatible with the bispectrum data, with $v_1=1$ excluded at $4.7\sigma$. This is due to the high amplitude of the baseline, squeezed, component, which is around two orders of magnitude higher than that of the standard squeezed template and that cannot be cancelled by the equilateral components without significantly increasing the equilateral non-Gaussian signal beyond the \textit{Planck} constraints, independently of the choice of the free parameters. This problem can be rectified by modifying the power spectrum parameters to reduce the amplitude of the baseline component, but still satisfying the power spectrum constraints. Such parameters are given by: $f_2 = 6$, $\alpha_B = 1$  $\alpha_H = 1.04$, $\beta_K = 3.97343$, $m^2 = -10^{-23}$, $\lambda = 1.2 \times 10^{-36}$ and $h_{\rm ds} = 0.00001712$. The baseline amplitude is reduced by a factor of around 50, allowing a sufficiently important cancellation from the equilateral component to match the \textit{Planck} constraints.


\section{Conclusions}
In this work we have shown that DHOST theories can be used to describe the inflationary epoch of the Universe. The models present several free parameters, that can be tuned to fit \textit{Planck} measurements for the power spectrum and the bispectrum. However, models that are compatible with the power spectrum data can be excluded by the bispectrum due to the large amplitude of a fixed squeezed component that cannot be cancelled by the additional bispectrum parameters. 
This work is based on ~\cite{dhost1,dhost2,dhost3}, in collaboration with P. Brax, W. Sohn and J.R. Fergusson.

\section*{Acknowledgements}
This work was supported by a UKRI Future Leaders Fellowship [Grant No. MR/V021974/2].

\end{document}